\def\aap{A\&A}%
\def\mnras{MNRAS}%
\def\pasp{PASP}%
\documentclass[traditabstract]{aa} % for the abstract without structuration 
\usepackage{longtable}
\usepackage{graphicx}
\usepackage{latexsym}
\usepackage{amssymb}
\usepackage{lscape}
\usepackage{txfonts}
\title{VANDELS consortium public database}
\author{L. Pentericci\inst{1}
\and R.J.McLure\inst{2}
\and P. Franzetti\inst{1}
\and B. Garilli\inst{1} 
and the VANDELS team}
\authorrunning{L.Pentericci et al.}
\titlerunning{VANDELS database}

\institute{ 
INAF-Osservatorio Astronomico di Roma , via Frascati 33, I-00078 Monteporzio Catone, Italy 
\and 
Institute for Astronomy, University of Edinburgh, Royal Observatory, Edinburgh, EH9 3HJ, UK 
\and 
INAF-Istituto di Astrofisica Spaziale e Fisica Cosmica Milano, via Bassini 15, I-20133, Milano, Italy 
}

\begin{document}
%\date{Received; accepted}

% \abstract{}{}{}{}{} 
% 5 {} token are mandatory
 
  \abstract
  % context heading (optional)
 {We provide access to the  current VANDELS data release (DR2)  through a dedicated database, accessible at the link  http://vandels.inaf.it/db.
The data release includes the spectra for all galaxies for which the scheduled integration time was completed during the first two observing seasons and the  spectra for those galaxies for which the scheduled integration time was 50\% complete at the end of season two (i.e. 20/40 hours and 40/80 hours). The total number of spectra released is 1362 (586 in CDFS and 776 in UDS).
Spectra are stored as multi-extension FITS files containing  the 1D extracted spectrum, the 2D linearly re-sampled spectrum, the 1D sky spectrum, the 1D noise estimate and  the image thumbnail of the object. 
}   

   \maketitle
%
%________________________________________________________________

\section{VANDELS}
VANDELS: a deep VIMOS survey of the CANDELS UDS and CDFS fields, is the latest ESO public spectroscopic survey.
The main targets of VANDELS are star-forming galaxies at redshift 2.4 $<$z $<$ 5.5, an epoch when the Universe had not yet reached 20\% of its current age, and massive passive galaxies in the range 1 $<$ z$ <$ 2.5. Details of target selection and survey strategy were presented in McLure et al. (2018).

VANDELS adopted a strategy of ultra-long exposure times, ranging from a minimum of 20 h to a maximum of 80 h per source, and is therefore the deepest-ever spectroscopic survey of the high-redshift Universe. Exploiting the red sensitivity of the refurbished VIMOS spectrograph, the survey has  obtained ultra-deep optical spectroscopy covering the wavelength range 
4800-10 000 \AA\ with a sufficiently high signal-to-noise ratio to investigate the astrophysics of high-redshift galaxy evolution via detailed absorption line studies of well-defined samples of high-redshift galaxies. The data were reduced using the full automated pipeline Easy-Life which is an updated version of the VIPGI system (Scodeggio et al. 2005). Details of  survey layout, observations and data reduction were presented in Pentericci et al. (2018).

The VANDELS raw data are immediately public and the collaboration is committed to periodical releases of the fully reduced and calibrated data through the ESO archive. In addition to spectra we  also provide  a catalog with the essential galaxy parameters, including spectroscopic redshifts and redshift quality flags measured by the collaboration, using dedicated tools that were developed and previously applied e.g. to the VUDS survey (Le F\`evre et al. 2015).
As an alternative to the ESO archive we now provide access to the data 
  through a dedicated consortium database, accessible at the link  http://vandels.inaf.it/db.
The main difference with the products available in the ESO archive is the format. In the new database 
spectra are stored as multi-extension FITS files containing the following:
\\
     $\bullet$  Primary: the 1D extracted spectrum
    \\
    $\bullet$ EXR2D: the 2D linearly resampled spectrum\\
     $\bullet$  SKY: the 1D sky spectrum\\
     $\bullet$  NOISE: the 1D noise estimate\\
     $\bullet$  EXR1D: a copy of the 1D extracted spectrum (to recover any edit that might be done on the Primary)\\
     $\bullet$  THUMB: the image thumbnail of the object\\
      $\bullet$ BLUE-CORR: the correction that was applied to the originally calibrated spectra, due to a systematic drop in flux at the very blue end of the spectra, compared to the available broad band photometry (see Pentericci et al. 2018 for more details).
    
Each extension can be viewed using standard tools like IRAF or IDL.  The full information stored in these files can be inspected as a
  whole using the pandora.ez software (Garilli et al. 2010) downloadable
  at this link; http://pandora.lambrate.inaf.it/docs/ez/quick-guide.html.

Currently the database contains the same data as in DR2, 
i.e. spectra obtained during the first two VANDELS observing seasons (ESO run numbers  194.A-2003(E-Q)), which ran  from Aug 2015-Jan 2016 and Aug 2016-Jan 2017, respectively. The data release includes the spectra for all galaxies for which the scheduled integration time was completed during the first two seasons. In addition, the data release also includes the spectra for those galaxies for which the scheduled integration time was 50\% complete at the end of season two (i.e. 20/40 hours and 40/80 hours). The total number of spectra released is 1362 (586 in CDFS and 776 in UDS).

%\begin{figure}[]
%\centering
%\includegraphics[width=9cm]{wedge.ps}
% \caption{   }
% \label{campione} %1
%\end{figure}

%\newpage
\end{document}